\documentclass[conference]{IEEEtran}
\usepackage[table,xcdraw]{xcolor}
\usepackage{multirow}
\usepackage{diagbox}
\ifCLASSINFOpdf
  \usepackage[pdftex]{graphicx}
\else
\fi
\ifCLASSOPTIONcompsoc
  \usepackage[caption=false,font=normalsize,labelfont=sf,textfont=sf]{subfig}
\else
  \usepackage[caption=false,font=footnotesize]{subfig}
\fi
\begin{document}
%
\title{A Method Based on Deep Learning for the Detection and Characterization of Cybersecurity Incidents in Internet of Things Devices
}

\author{\IEEEauthorblockN{Jhon Alexánder Parra-Jiménez}
\IEEEauthorblockA{Facultad de Minas\\
Universidad Nacional de Colombia\\
Email: jparraj@unal.edu.co}
\and
\IEEEauthorblockN{Sergio Armando Gutiérrez-Betancur}
\IEEEauthorblockA{Facultad de Ingenierías\\
Universidad Autónoma Latinoamericana\\
Email: sergio.gutierrezbe@unaula.edu.co}
\and
\IEEEauthorblockN{John Willian Branch-Bedoya}
\IEEEauthorblockA{Facultad de Minas\\
Universidad Nacional de Colombia\\
Email: jwbranch@unaula.edu.co}}


%


\maketitle

\begin{abstract}
Given the increased growing of Internet of Things networks and their presence in critical aspects of human activities, the security of devices connected to these networks becomes critical. Machine Learning approaches are becoming prominent as enablers for security solutions in computer networks due to its capacity to process traffic information in order to detect abnormal patterns which might represent attacks targeting infrastructures. In this paper, we propose to leverage Convolutional and Recurrent Neural Networks, two artifacts that have been successfully used in contexts such as image processing for pattern recognition, for the development of a security solution to be used in the context of Internet of Things. Our results show that this approach, when evaluated with a state-of-the-art data set, achieves around 99\% of accuracy in the binary classification of attacks (i.e. normal traffic vs attack traffic) and 96\% for multiclass classification (recognition of different types of attacks) accuracy. These results outperform proposals available in literature, showing a promising landscape for developing security solutions for IoT infrastructures. 
\end{abstract}


%
\IEEEpeerreviewmaketitle

\section{Introduction}

The technological revolution observed during the last 50 years has led to a hyper-connected society which strongly depends on large volumes of information stored in devices connected to Internet \cite{parviainen2017tackling}. The interconnection among many of these devices introduces the Internet of Things (IoT) paradigm, which is basically defined as a set of devices, usually equipped with sensor capacities and connected through a public or private communications network enabling data exchange among these devices \cite{khan2018iot}. It is estimated that for 2025, these networks will comprise up to 21.5 billions of devices \cite{statista2020}, with an approximated volume of up to 175 Zetabytes of data exchanged by devices \cite{rydning2018digitization}.

Given the characteristics of IoT devices, including limited processing and storage capabilities, they are prone to a large number of vulnerabilities. Considering the amount of information they produce and process, and the involvement of IoT devices in critical infrastructures, their security becomes a very relevant research topic. In this work, we present our results in the design and development of a method for the early detection and characterization of cybersecurity incidents in the context of IoT networks. Hence we propose a solution leveraging advanced Machine Learning techniques, which shows to be highly effective in comparison to state-of-the-art approaches.

The remaining of this paper is organized as follows. Section \ref{related} presents a discussion on the main threats for IoT networks, and the state-of-the-art proposals for the protection of devices in these networks. Section \ref{proposal} presents a general description of the proposed model. Section \ref{experiments} presents the experiments performed to evaluate the proposed model, and finally Section \ref{conclusions} presents the conclusions of our work.

\section {Related Work}
\label{related}

The accelerated growing in the deployment and implementation of IoT devices in recent years has increased the interest on these infrastructures, and specially on their security. In this section, we present an overview of recent works that have addressed the challenge of protecting IoT systems.

From a Systematic Literature Review process, we have found that Machine Learning (ML) techniques are widely used in the context of IoT devices since they overcome different limitations of traditional approaches \cite{hussain2020machine}.  Recent works \cite{yugha2020survey,hassija2019survey,haddadpajouh2019survey,miloslavskaya2019internet,anand2020iot, baig2020averaged, paudel2019detecting, galeano2020detection} show specially a growing interest in the application of Deep Learning techniques to address IoT security challenges.


Ujjan et al. \cite{ujjan2020towards} use a Stacked AutoEncoder for classification of outlier records. This is an approach focused on detecting DoS attacks. Authors also use multiple sparse autoencoders with three layers each, and a Softmax function for the detection of outliers. This enables the integration with other models in order to classify the attacks by only filtering potential outlier requests.

Roopak et al. \cite{roopak2019deep} focus on the comparison of the performance of several different neural networks architectures against traditional ML models such as Support Vector Machines (SVM), Naive Bayes and Random Forests. The compared architectures include Convolutional Neural Networks (CNN), Multilayer Neural Networks, Long short-Term Memory (LSTM) Recurrent Neural Networks, and a mix of CNN and LSTM Neural Networks. The evaluation of these proposals shows that the best performance is achieved with the mix of CNN and LSTM, with an accuracy up to 97\%.

Finally, Hussain et al. \cite{hussain2020iot} present a non conventional technique. They use a CNN which performs classification by mapping data packets and flow information onto a image representation. This approach achieves results with noticeable improvements in comparison with similar models. It achieves up to 99.9\% of accuracy. 

\section{Proposed Model}
\label{proposal}

\subsection{Background}

According to the information collected from related works, we decided to base our proposed solution on the idea of Hussain et al. \cite{hussain2020machine}. Hence, we leverage the features and potential of CNN when applied to image processing and discovering patterns potentially hidden within them. Specifically, the procedure is:

\begin{itemize}

\item Feature engineering is performed in order to include only those variables relevant for the attack classification.
\item Selected features are normalized, so that their values lie in the [0-1] interval, using their minimum and maximum values.
\item With $n$ variables, the same amount of $n$ registers is selected in order to build the first channel for a RGB image, for each category of the response variable. This step is performed twice for the remaining records. Hence, each image becomes composed of $3xn$ records, producing a vector of $n x n x 3$ positions.
\end{itemize}

These steps are repeated until covering all the available records.


With the data set created in this way, it is possible to train models such as CNNs. For instance, Hussain et al. \cite{hussain2020machine} trained a residual neural network \textit{resnet18} (A variation of traditional CNNs), which uses jumps between layers to avoid the descendent gradient problem and better preserving the information learnt across the time.

\subsection{Proposed approach}

Our proposal is based on converting data from network packets into images in order to take advantage of famous CNN architectures such as Xception \cite{chollet2017xception}, Inception \cite{szegedy2016rethinking} and Resnet \cite{he2016identity} for feature extraction from the processed images. This architecture is selected as reference point, similar to the model used by Hussain et al. \cite{hussain2020machine}. The reason to use these architectures are: a) Their performance, since the three mentioned architectures are found among the 10 best regarding accuracy in the validation set of ImageNet; b) their size on disk, which is an important factor if the system is going to be implemented within an IoT system. The size on disk of these models is not larger than 100MB, which indicates that these modes have less parameters than other architectures. The comparison of the parameters of the chosen models vs different modes is show on Table \ref{table:parameters}.

\begin{table}[!t]
\renewcommand{\arraystretch}{1.3}
\label{table:parameters}
\caption{Comparison among the parameters of the chosen models and other models}
\centering
\begin{tabular}{|l|l|l|}
\cline{1-3}
\bfseries Model & \bfseries Size on Disk  & \bfseries Parameters  \\ \hline
\cellcolor[HTML]{9B9B9B} Xception & \cellcolor[HTML]{9B9B9B} 88 MB & \cellcolor[HTML]{9B9B9B} 22,910,480 \\ \cline{1-3}
\cellcolor[HTML]{9B9B9B} InceptionV3 & \cellcolor[HTML]{9B9B9B} 92 MB & \cellcolor[HTML]{9B9B9B} 23,851,784  \\ \cline{1-3}
\cellcolor[HTML]{9B9B9B} ResNet50 & \cellcolor[HTML]{9B9B9B} 98 MB & \cellcolor[HTML]{9B9B9B} 25,636,712 \\ \cline{1-3}
NASNetLarge & 343 MB & 88,949,818  \\ \cline{1-3}
InceptionResNetV2 & 215 MB & 55,873,736  \\ \cline{1-3}
ResNet152V2 & 232 MB & 60,380,648 \\ \cline{1-3}
ResNet101V2 & 171 MB & 44,675,560 \\ \cline{1-3}
DenseNet201 & 80 MB & 20,242,984 \\ \cline{1-3}
\end{tabular}
\end{table}

According to authors such as Roopak et al. \cite{roopak2020intrusion}, a configuration which might deliver interesting results for the detection of DoS attacks is the combination of CNN with Recurrent Neural Networks. The conjecture regarding this configuration is that whereas CNN provide capabilities for the extraction of hidden patterns in images, and the mapping of these parameters onto a compact dimensional space, Recurrent Neural Networks can analyze sequences of data containing information about the time where a given event occurred (e.g. an attack, or the sending/receiving of information), and therefore, to capture the nature of the patterns that might arise during DoS attacks. Hence, we base our proposal in a combination of these two architectures.

\subsection{Definition}

\begin{figure*}[t!]
    \centering
    \includegraphics[scale=0.4]{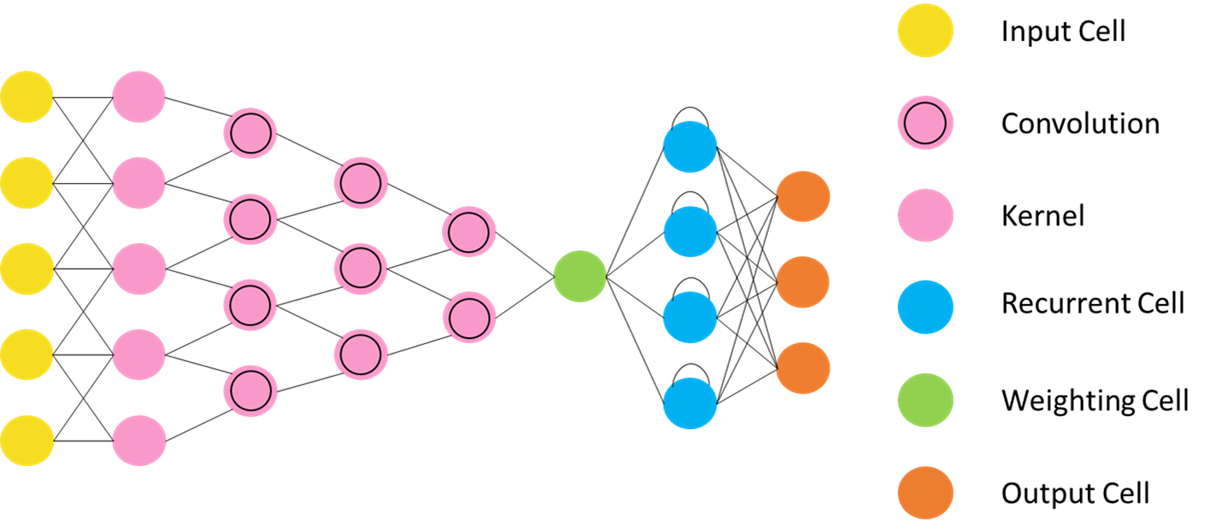}
    \caption{Base architecture of the models to train}
    \label{fig:Arch}
\end{figure*}

The convolutional layers will serve as feature extractors from the images where patterns will be identified, and later decoded in a vector of numbers. Recurrent layers will be fed with this vector of numbers representing the relevant features of each image, and then a recurrent cell type LSTM will be updated, which will store relevant information of event sequence. An event (either benign or malign) is defined as the sending or receiving of information in the IoT system. Figure \ref{fig:Arch} presents the base architecture of the proposed solution which will be used for the models to be trained.

In this architecture, the convolutional and recurrent components previously described can be observed. The first layer is composed by neurons on charge of receiving the information stored in the images. Each input layer will directly depend on the architecture used for feature extraction. Thus, for Xception 71x71 images are required, for Inception, images should be 75x75 and for Resnet they should be 32x32.

Once the images are received in the neural network, they are processed through the particular architecture of the considered models. For instance, Xception has almost 23 million of parameters, Inception has around 24 millions of parameters and ResNet around 25.5 million of parameters. Next, we introduce some of the main features of each model:

\begin{itemize}
    \item Xception: This model is based on Inception. The main difference is that this architecture modifies the inception blocks and replaces them with separated convolutional blocks of 1x1 and 3x3 to extract features \cite{chollet2017xception}. The main contribution of this architecture is the introduction of CNNs with convolutional separable layers.
    \item Inception V3: This model is built on the idea of existing correlations on each image in such a way there is an inception block which simultaneously extracts features by applying 1x1,3x3 and 5x5 convolutions which are later concatenated. Its main contribution is the use of batch normalization \cite{szegedy2016rethinking}.
    \item ResNet 5.0: Different to the previous architectures, these networks use to be deeper, with a difference which is jumps between connected layers. This helps to solve the descendent gradient and the training problems by connecting and providing results coming from different layers \cite{he2016identity}. Its main contribution is the generation of deeper CNNs without compromising the generality of the model.
\end{itemize}

After this, data are sequentially processed in 64 LSTM neurons, and later, data are further fed into 3 neurons which generate the final result of the class that the image belongs to.

\section{Experiments and Results}
\label{experiments}

In this section, we present in detail the implementation of a proposal of a DoS attack detection model in IoT networks. This section is organized in 3 subsections: First, we introduce the data set containing samples and traces of DoS attacks in the context of an IoT network and the preprocessing and analysis we performed on it. Second, we implement the different neural network architectures considering the selected data set. Third, we identify from the literature some related works that can be useful to compare the results obtained with our approach.

\subsection{Datasets of DoS attacks in IoT}

For the selection of the data set, we performed a literature survey aiming at identifying data sets specific for the context of IoT. From this survey, we identified Bot-IoT data set as one of the most relevant available in literature. Bot-IoT is a data set specific for IoT developed by researchers at UNSW in Canberra which includes samples of DoS attacks against different kind of devices present in an IoT network \cite{koroniotis2019towards}. Bot-IoT contains more than 73M of traffic records, including 29 variables. Previous to our experiments, we preprocessed the files conforming this data set in a server running Linux kernel 4.19.112, equipped with 36GB of RAM, 40 Intel Xeon CPUs running at 2.3Ghz and 8 TPU cores. From the raw data set, we identified 6 empty variables. The data set files were injected into a SPARK instance in order to be able to handle it. 


In Bot-IoT data set, most of the records are associated to DoS and DDoS attacks. These classes represent 97.5\% of the records. Hence, it is important to balance the data. From the original classes contained in the data set, we excluded the Stealing category, and we merged the Normal and Reconnaissance categories in a single category named as "Others". We used controlled sub-sampling for further balancing of the data set. That is, we considered the most representative classes and we selected some portions of data but preserving the sequencing of the records. This introduces an important challenge for us, consisting in selecting a representative sample that preserves the sequence of events. To obtain a balanced data set, we would require taking at least the 100\% of the records contained in the "Others" category, 5.54\% of the DoS records and 4.75\% of the records in the DDoS class. However, since we have the requirement of preserving the sequential nature of the records in order to preserve their temporal aspect \cite{yin2017deep}, we had to take a larger sample of the records in the "Others" category. Table \ref{table:finalData} presents the final composition of our data set.

\begin{table}[t!]
\renewcommand{\arraystretch}{1.3}
\caption{Final composition of the evaluation data set}
\label{table:finalData}
\centering
\begin{tabular}{|l|l|l|} \hline
\bfseries Class & \bfseries Records & \bfseries Percentage \\ \hline
 DDoS & 4816344 & 44.71\% \\ \hline
 DoS & 4125279 & 38.29\%  \\ \hline
 Others & 1831182 & 17.00\%  \\ \hline
 \end{tabular}
\end{table}


\subsection{Model for the detection and classification of DoS attacks in IoT}

In this section, we describe in detail our method for detection and classification of DoS attacks in the context of IoT networks. Basically, our method consists of transforming the tabular data contained in the data set into RGB images with three channels. With these images, we train three neural networks models based on convolutional and recurrent architectures. For this transformation, we take 16 features and 16 records from the data set. Thus, we build 16 x 16 images to form each of the channels of the image. The values of the features are normalized and later scaled so that they take values between 0 and 255. Equation \ref{eq:minMax} presents the scaling function used for this step.

\begin{equation}
\label{eq:minMax}
x\prime_{i}=\frac{x_i - min(x_i)_{i}}{max(x_{i})_{i} - min(x_{i})_{i}}
\end{equation}

\begin{figure}[t!]
    \centering
    \includegraphics[scale=0.5]{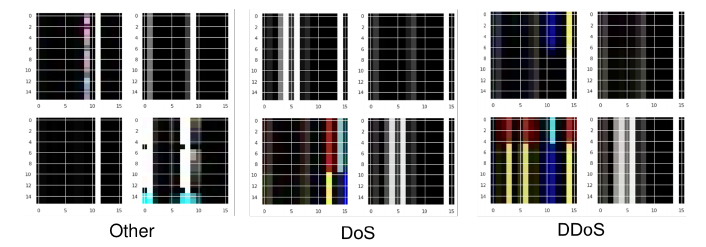}
    \caption{Examples of the generated RGB images}
    \label{fig:examples}
\end{figure}

The transformed values are later converted into a $16x16x3$ arrays for each class. That is, we separate the records of each class (DoS, DDoS and others) and we generate the images of three channels with 48 records where the first channel corresponds to the first batch of 16 records, the second channel corresponds to the following 16 records and the third channel corresponds to the final 16 records. This process is repeated until covering the records available for each class. From the data set, we obtained 100340 images of DDoS class, 85943 images of DoS class and 38149 of the others class. Figure \ref{fig:examples} presents some examples of the obtained images for each class. Since the data set is not completely balanced, we modified the relative weights of each record for the model training in order to avoid any possible overfitting. Equation \ref{eq:weights} shows how these weights were calculated for each sample of each class (DoS, DDoS and Others).

\begin{equation}
\label{eq:weights}
RelativeWeight_i = \frac{TotalRecords}{AmountOfRecords_i}
\end{equation}

Thus, the weights for each class are 2.2367 for DDoS, 2.6114 for DoS and 5.8830 for Others. This means that a wrong classification for the Others class is 2.6302 times more expensive than a DDoS and 2.2528 times more expensive than a DoS. Hence, the potential loss of each class is the same and the model optimizes the loss function considering these weights.

Since the images have been generated with 16x16 matrices, we performed a further pre-processing that adjusts the images to the pixel resolution of each of the convolutional neural networks we used (Xception, Inception, ResNet50) by using bi-linear interpolation of each frame.

For the model training, we used a learning rate of 0.0001 during 50 epochs, with a batch size of 64 images in order to maximize the right classification rate and the F1 score on the validation set. The evaluation considered the metrics of Accuracy, Precision, Recall and F1-Score. These metrics are defined according to the formulas indicated in \cite{roopak2020intrusion}.






Table \ref{table:results} presents the results of the best iteration for each trained model. From these results, it is possible to see that the best model is \textbf{Inception} as feature extractor \textbf{combined with a LSTM layer} of 64 nodes since it achieves the best values for the metrics on the validation set. Also, it tends to present a stable performance in both of the sets, similar to the results observed with Resnet50. Regarding Xception, the results show to have an error 2\% higher on the training set, in comparison with the validation set. From these results, we have decided to continue using the Inception model for the comparison with related work.

\begin{table*}[t!]
\renewcommand{\arraystretch}{1.3}
\caption{Results of the training phase for the considered models}
\label{table:results}
\centering
\begin{tabular}{|l|ll|ll|}
\hline
\multicolumn{1}{|c|}{\multirow{2}{*}{\bfseries Model}} & \multicolumn{2}{l|}{\bfseries Training}                          & \multicolumn{2}{l|}{\bfseries Validation}                        \\ \cline{2-5} 
\multicolumn{1}{|c|}{}                       & \multicolumn{1}{l|}{\bfseries Correct Classification} & \bfseries F1 Score & \multicolumn{1}{l|}{\bfseries Correct Classification} & \bfseries F1 Score \\ \hline
Xception + LSTM                              & \multicolumn{1}{l|}{0.948}                  & 0.948    & \multicolumn{1}{l|}{0.9669}                 & 0.9665   \\ \hline
Inception + LSTM                             & \multicolumn{1}{l|}{0.9629}                 & 0.9629   & \multicolumn{1}{l|}{0.9696}                 & 0.9693   \\ \hline
Resnet50 + LSTM                              & \multicolumn{1}{l|}{0.9625}                 & 0.9625   & \multicolumn{1}{l|}{0.9628}                 & 0.9626   \\ \hline
\end{tabular}
\end{table*}

In the remaining experiments, we assess the results obtained with the model based in Inception + LSTM in order to observe its specific performance. Table \ref{fig:confusion} presents the confusion matrix of the model when evaluated with the validation data set. From this matrix, it is possible to observe that most of the incorrect classifications are in the DDoS and DoS classes. However, we have determined that the correct classification of a record as either attack or in the "Others" class (regardless the specific class of attack) is around 99.8\% with a F1-Score of 0.999. This indicates the model can detect with good precision and accuracy if a record represents an attack to the IoT system. 

\begin{table}[t!]
\renewcommand{\arraystretch}{1.3}
\caption{Confusion matrix for the model with Validation Set}
\label{fig:confusion}
\centering
\begin{tabular}{lllll}
\cline{3-5}
                          & \multicolumn{1}{l|}{DDoS}   & \multicolumn{1}{l|}{\cellcolor[HTML]{3531FF}{\color[HTML]{FFFFFF} 19119}} & \multicolumn{1}{l|}{\cellcolor[HTML]{68CBD0}870}                          & \multicolumn{1}{l|}{\cellcolor[HTML]{38FFF8}30}   \\ \cline{3-5} 
                          & \multicolumn{1}{l|}{DoS}    & \multicolumn{1}{l|}{\cellcolor[HTML]{68CBD0}419}                          & \multicolumn{1}{l|}{\cellcolor[HTML]{3166FF}{\color[HTML]{FFFFFF} 16647}} & \multicolumn{1}{l|}{\cellcolor[HTML]{38FFF8}27}   \\ \cline{3-5} 
                          & \multicolumn{1}{l|}{Others} & \multicolumn{1}{l|}{\cellcolor[HTML]{96FFFB}10}                           & \multicolumn{1}{l|}{\cellcolor[HTML]{96FFFB}2}                            & \multicolumn{1}{l|}{\cellcolor[HTML]{34CDF9}6630} \\ \cline{3-5} 
\multirow{-4}{*}{\rotatebox{90}{Current}} &                             & DDoS                                                                      & DoS                                                                       & Others                                            \\
                          & \multicolumn{4}{c}{Predicted}                                                                                                                                                                                                          
\end{tabular}
\end{table}

\begin{table}[t!]
\renewcommand{\arraystretch}{1.3}
\caption{Complete results of the assessment of the proposed model}
\label{table:completeresults}
\centering
\begin{tabular}{|l|l|l|l|l|l|}
\hline
\backslashbox{Class}{Metric} & Precision & Recall & F1-Score & Accuracy \\ \hline
DDoS  & 0.98 & 0.96 & 0.97 & 0.955  \\ \hline
DoS   & 0.95 & 0.97 & 0.96 & 0.974  \\ \hline
Other & 0.99 & 1 & 0.99 & 0.998  \\ \hline
\end{tabular}
\end{table}

Table \ref{table:completeresults} presents the complete results of precision and exhaustiveness of the model, and also the weighted rate of classification. In the next section, we present the results of comparing the proposed model against other proposals found in literature.

\subsection{Comparison with Related Work}

In this section, we present a comparison of our proposal with different works available in literature which address the problem of detection and classification of attacks in the context of IoT Networks. Since we considered the Bot-IoT dataset, we focus our comparison on works that also used this same data set. We identified from our literature review five papers addressing the same problem based on the same data set.

Susilo and Sari \cite{susilo2020intrusion} discuss different machine learning and deep learning techniques to improve the performance of cybersecurity techniques in IoT networks. They train different models for the detection and classification of DoS. Authors do not specify if they train their model using the whole data set or a sample of it. However, they find that the model with the best performance is a convolutional network, achieving a classification rate of 0.913 in the categorization of the attack with each one of the categories present in the data set.

Pokhrel et al \cite{pokhrel2021iot} trained several models such as KNN, Naive Bayes and a Perceptron Multilayer Neural Network. They used one of the 74 files of the data set, which contained both attack and normal traffic samples. Given the unbalance present in the data set, the authors used the SMOTE algorithm in order to create synthetic records. The model that achieved the best performance was KNN with an correct attack classification rate of 0.921.

Ibitoye et al \cite{ibitoye2019analyzing} present a different approach where they compare two neural networks architectures for detection and classification of DoS attacks. Authors use the reduced version of the Bot-IoT data set which contains 5\% of the original data, obtained randomly. This sample contained only the ten more relevant variables following the work by Koroniotis et al \cite{koroniotis2019towards}. Then, they train a Feedforward Neural Network (FNN) and a Spiking Neural Network (SNN). They compare these two architectures and observe that FNN outperforms SNN. It presents an accuracy and F1-Score of 0.95 while SNN presents accuracy and F1-Score of 0.91, both cases assessed with the validation set. However, when adversary records (That is, records aiming at confusing the model) are introduced, SNN shows to be more resilient and tends to show a better behavior in comparison to FNN. 

Shafiq et al \cite{shafiq2020iot} propose a method for feature selection so that those features with the highest contribution to the detection and classification of attacks become selected. The method proposed by these authors uses the correlation between variables and estimates iteratively which of the variables should be either included or excluded in the final model. Hence, they reduce from 39 to 7 the number of features to consider, which leads to metric values up to 95\% in the performance metrics considered for four algorithms: SVM, Naive Bayes, C4.5 tree and Random Forest. Authors conclude that their proposed method shows promising advances in feature selection while the model performance does not become adversely affected by removing features.

Finally, Injadat et al \cite{injadat2020detecting} present a model based on a decision tree for the detection of attacks. They use a reduced version of the set composed by 5\% of the records randomly selected. After this selection, data are normalized and SMOTE algorithm is applied in order to balance the data set. Then, they use a Bayessian Optimization Gaussian Process (BO-GP) in order to optimize the hyperparameters of the decision tree. With the optimized parameters the model is trained. The accuracy obtained is 0.99 and the F1-Score is 1. These results are contrasted with a standard decision tree and a SVM, and the results obtained demonstrated the efficacy of the BO-GP for the feature selection. 

\begin{figure*}[ht]
    \centering
    \subfloat[Multiclass]{%
       \includegraphics[width=0.45\linewidth]{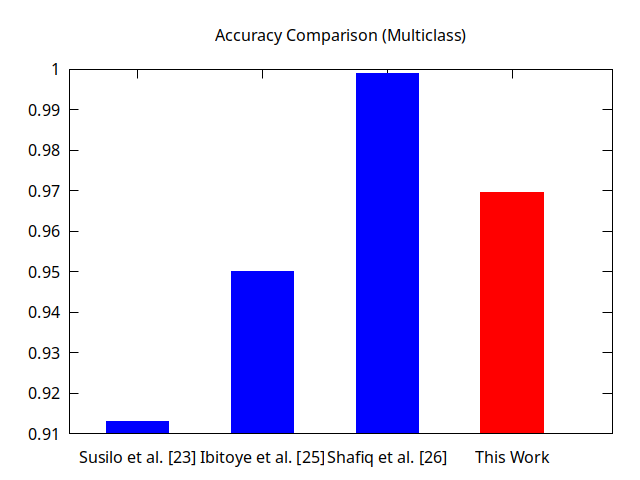}}
    \hfill
  \subfloat[Binary]{%
        \includegraphics[width=0.45\linewidth]{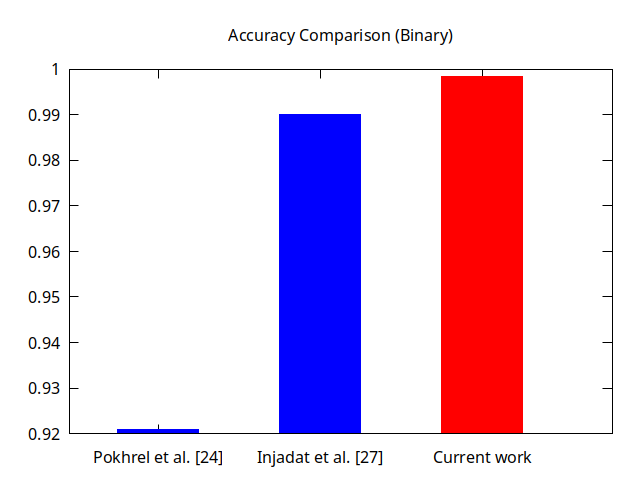}}
    \\    
    \caption{Comparison of the Accuracy metric in different scenarios}
    \label{fig:comparison}
\end{figure*}

Figure \ref{fig:comparison} presents the consolidated view in terms of the correct classification rate. These results show how the performance of our model is similar to the performance exhibited by other works available in the literature. Actually, the performance of our model is the highest in the problem of binary classification whereas it is the second best performance in the multiclass classification. Here it is important to clarify this is a preliminary comparison, given the fact that related works have used a subset of the original data. Also, in some of the works, techniques such as SMOTE were used to balance the data sets. Nevertheless, from this preliminary comparison, it is possible to see that our proposal presents a good performance in the task of detecting and classifying attacks in the context of IoT networks. It is also a novel proposal in the sense of using the transformation of tabular data into images in order to take advantage of the power of convolutional networks, and including recurrent cells to consider the temporal component which is inherent to the intrusion detection problem. Hence, we consider that our proposed approach introduces an interesting line of research to be explored in the future.


\section{Conclusions and Future Work}
\label{conclusions}

In the current context, the attack with the highest impact against IoT systems is DoS, since this is the attack which is reported the most in the literature in recent years. Methods for detection and classification of attacks are becoming and important research topic. Most of the data sets available for the training of models for attack detection are tailored towards general purpose cybernetic systems. From the experiments we performed in our assessment, we could observe that a model based on an Inception Deep Neural Network combined with a recurrent LSTM layer offered a great capacity for the detection and classification of DoS attacks in the context of IoT. The values for the metrics obtained in these experiments resulted to be superior to those obtained in related work. Our proposal has a great potential to be yet improved by incorporating larger feature extractors which might benefit the specific attack classification. That is, the multiclass problem. Our model showed in the experiments that it can detect DoS attacks with an accuracy rate of 99.84\%. This is a promissory result for the exploration of additional attack classes, as future work. An aspect yet to be better assessed is the velocity of detection. Finally, another important further work that needs to be addressed is evaluating the deployment of our proposal in a real system, perhaps integrated with a firewall or network control element (e.g. A Software Defined Network Controller).

\section*{Acknowledgment}
This paper has been supported by the Ibero-American Science and Technology Program CYTED (Project: 519RT0580).

\bibliographystyle{IEEEtran}
\bibliography{IEEEabrv,biblio}

\end{document}